\documentclass[12pt]{article}

\input epsf
\def\beq{\begin{eqnarray}}
\def\eeq{\end{eqnarray}}

\def\lsim{\mathrel{\rlap{\lower3pt\hbox{\hskip0pt$\sim$}}
    \raise1pt\hbox{$<$}}}         
\def\gsim{\mathrel{\rlap{\lower4pt\hbox{\hskip1pt$\sim$}}
    \raise1pt\hbox{$>$}}}         

\begin{document}


\vskip 1cm
\begin{center}
{\Large \bf  A Vacuum Accumulation Solution to the Strong CP Problem}

\vskip 1cm {Gia Dvali\footnote{\it  email:  dvali@physics.nyu.edu}}

\vskip 1cm
{\it Center for Cosmology and Particle Physics, Department of Physics, New York University, New York, NY 10003}\\
\end{center}

\vspace{0.9cm}
We suggest a solution to the strong CP problem in which there are no axions involved. 
The superselection rule of the $\theta$-vacua 
is dynamically lifted in such a way that an infinite number of vacua are accumulated within the phenomenologically acceptable range of  $\theta < 10^{-9}$, whereas only a measure-zero set of vacua remains outside of
this interval.  The general prediction is the existence of  membranes to which the standard model  gauge fields are coupled. These branes may be light enough for being produced at the particle accelerators in form of the resonances with a characteristic membrane spectrum.  

\vspace{0.1in}

\newpage

\section{Introduction}

The strong CP problem\cite{theta} can be expressed as an inexplicable smallness  of the CP-violating $\theta$-parameter in QCD Lagrangian
\begin{equation}
\label{aff}
\theta \, {\rm Tr } G_{\mu\nu}\tilde G^{\mu\nu},
\end{equation}
where $G_{\mu\nu}$  and $\tilde{G}^{\mu\nu} \, = \, \epsilon^{\mu\nu\rho\kappa}G_{\rho\kappa}$ 
are the QCD gauge field strength and its dual respectively.   
A non-zero $\theta$ implies a non-zero expectation value of the dual gauge field strength,
$\langle$Tr$G\tilde G\rangle\, \neq 0\, $, which would lead to the observable CP 
violation, unless $\theta$ is tiny, $\theta < 10^{-9}$\cite{review}. 

 The strong CP problem is the problem of the vacuum superselection. The parameter $\theta$ scans 
 a continuum of the vacuum states that have different expectation values of the operator Tr $G\tilde{G}$. 
 These vacua satisfy the superselection rule, which forbids any transition between the vacua with 
 different values of $\theta$. 
 Because of this superselection rule, there is no {\it a priory} reason 
to give any preference to the vacua with small $\theta$ (small Tr $G\tilde G$). 

 In this paper we shall propose a new approach to this problem, in  which the $\theta$ vacua become
 rearranged in such a way that a vacuum with an
 acceptably small $\theta$ becomes `infinitely preferred'  relative to the other vacua. 
In our treatment, the superselection rule gets partially lifted, but in such a way that 
the infinitelly many vacua accumulate inside the region $\theta \ll 10^{-9}$, whereas only a measure-zero set of vacua remain outside this interval. 

   Our solution is based on the following key facts.  First, because a non-zero $\theta$ implies a nonzero 
  vacuum expectation value  $\langle {\rm Tr} G\tilde G \rangle$,
the explanation of the small $\theta$ reduces to the explanation of the small value 
of  Tr$G\tilde G$. Note that  the latter gauge-invariant can be rewritten as a dual four-form field strength
 of a Chern-Simons three-form in the following way (below everywhere we shall work in units of the CQD scale) 
 \begin{eqnarray}
\label{apsi}
  {g^2\over 8\pi^2} {\rm Tr} G\tilde{G} \, = \, F\, \equiv \,  F_{\alpha\beta\gamma\delta}
\epsilon^{\alpha\beta\gamma\delta}, 
 \end{eqnarray}
where
 \begin{eqnarray}
\label{gluonfourform}
 F_{\alpha\beta\gamma\delta} \,= \partial_{[\alpha}C_{\beta\gamma\delta]}. 
 \end{eqnarray}
$C_{\alpha\beta\gamma}$ is a Chern-Simons three-form, which in terms of the gluon fields can be written as
\begin{equation}
\label{qcdform}
C_{\alpha\beta\gamma} \, = \,  {g^2\over 8\pi^2}\, {\rm Tr} \left (A_{[\alpha}A_{\beta}A_{\gamma]}\,  - {3 \over 2}
A_{[\alpha}\partial_{\beta}A_{\gamma]}\right ).
\end{equation}
Here $g$ is the QCD gauge coupling, $A_{\alpha} \, = \, A^a_{\alpha}T^a$ is the gluon gauge field matrix, and $T^a$ are the generators of the $SU(3)$ group. 

The above parametrization is not just a formality, but has a physical meaning. 
It is known\cite{qcdform}  that in low energy QCD the Chern-Simons three-form behaves as a 
{\it massless} gauge field, and the corresponding four-form field strength $F$ can assume an arbitrary constant value.  Hence, the strong CP problem can be simply understood as the problem of an
{\it arbitrary} constant four-form electric field.  Any dynamical solution that would explain why such a field
must be unobservably small, would also automatically solve the strong CP problem. 

 For example, as explained in \cite{axigauge}, the celebrated Peccei-Quinn solution \cite{pq} solves the strong CP problem by putting the three-form gauge theory 
into the Higgs phase.  This happens because the three-form gauge field becomes massive by `eating up' a would be  massless  axion\cite{axion}, and acquires a propagating longitudinal degree of freedom.  
This effect is analogous to an ordinary Higgs effect in which the photon acquires a longitudinal polarization
by eating up a Goldstone boson. 
The generation of the axion mass from the QCD instantons can be reformulated in this language as a three-form 
Higgs effect.  As a result of this effect,  the four-form electric  field is {\it screened}\cite{gigamisha}, and the vacuum is 
automatically CP-conserving.  Thus, the three-form language gives a very simple explanation 
to the fact \cite{vw} that the minimum of the axion potential is always at $\theta \, = \, 0$\cite{axigauge}. 
 
 In the present paper, we shall present an alternative solution to the strong CP problem. 
 We shall attempt to explain why the QCD four-form electric field (and thus $\theta$) is small by employing 
 the vacuum-accumulation mechanism of \cite{giaalex, attractors}. This mechanism is ready made for 
 the three-form gauge theories and provides a general framework  in which the small value of any  parameter, that is determined by a four-form electric field, can  become  an accumulation point of the infinite number of vacua. 
 
 Applied to the strong CP problem our strategy can be outlined as follows.  We shall lift the
 superselection rule of the $\theta$-vacua by postulating the existence of  2-branes  that source the 
 QCD Chern-Simons three-form $C$.  These branes substitute the axion field $a$ of the Peccei-Quinn solution in the following way.  In the Peccei-Quinn scenario the QCD Chern-Simons three-form
$C_{\alpha\beta\gamma}$  is sourced by the topological axionic  current 
$\epsilon_{\alpha\beta\gamma\mu}\partial^{\mu} a$.  This sourcing is the reason for screening 
the expectation value of $F$.  In our treatment this topological current gets 
replaced by a world-volume history of a dynamical two-dimensional surface, a 2-brane
\vskip 0.7 cm
\begin{equation}
\label{replace}
\epsilon^{\alpha\beta\gamma\mu}\partial_{\mu} a(x) ~~~ \rightarrow ~~~ 
\int dY^{\alpha\beta\gamma}\delta^4(x -Y),
\end{equation} 
\vskip 0.7 cm
 where  $Y^{\alpha}$-s are the target space coordinates of the brane. 
 In this way the massless scalar that screens $F$ gets replaced by a massive extended object, 
 that affects  the expectation value of $F$ in a different way.    
 The precise origin of these branes
 is unimportant for us. The only nontrivial assumption is that the 
 branes are  CP-odd.  The transition between the subset of vacua then becomes possible 
 quantum-mechanically via the 2-brane nucleation.  The branes in this picture play the role of the 
 domain walls that separate vacua with different values of $\theta$ (and $F$). Although at low energies the real  time transitions are extremely rare, this does not concern us, since we are mostly preoccupied 
 with the resulting vacuum statistics.  This statistics turns out to be pretty profound, because of the CP-odd nature of branes.
 
  By parity (P) and CP symmetries  the charge of the branes with respect to $C_{\alpha\beta\gamma}$ must also
 be parity-odd and is determined by the value of the parity-odd four-form electric field (\ref{gluonfourform}).  Thus, in vacua with 
 smaller electric field $F$, the sourcing is correspondingly weaker, and entirely diminishes in the
vacuum with  $F=0$ ($\theta = 0$).  In this way, the brane charge is set by $\theta$, and hence the step by which  $\theta$ changes from vacuum to vacuum is set by $\theta$ too. 
 This fact guarantees that the vacuum with  $\theta = F = 0$ is the maximally preferred  one. That is, essentially all 
of the infinite number of (quantum-mechanically-connected) vacua have arbitrarily small values of the $\theta$ parameter, and only a 
measure zero fraction has an observably-large CP violation.  

 Summarizing shortly,  due to the lift of the superselection rule the $\theta$-vacua get split in 
 discrete sets of the vacuum states, that we shall refer to as the {\it vacuum families}.  Each family contains an infinite number of vacua. 
The defining property of a given family is that  all its member vacua can be connected to one another 
by a quantum-mechanical tunneling, whereas the transition between the different families is forbidden. 
Within each family the number of vacua  as a function of the $\theta$-parameter
diverges for $\theta \rightarrow 0$ as
\vskip 0.7 cm
\begin{equation}
\label{nvac}
n_{\theta} \sim \theta^{-k} ~~~{\rm or} ~~~n_{\theta} \sim {\rm ln}( \theta^{-1}),
\end{equation}
\vskip 0.7 cm
where $k > 0$.  Thus, $\theta \, = \, 0$ is the vacuum accumulation point. Such points 
in the space of vacua where called the {\it vacuum  attractors} in \cite{giaalex, attractors}.
 
  This is the essence of the solution, which we shall discuss in details below. 
 
The general prediction of the above scenario is the existence of branes that source the Chern-Simons
forms of the Standard Model gauge fields.  The tension of these branes can be sufficiently low 
to be produced in particle collisions at LHC in form of the resonances with spacing and multiplicity characteristic to the brane spectrum.

\section{Three-Form Overview}
 
For a massless three-form field the lowest order parity-invariant Lagrangian
has the following form
\begin{equation}
\label{Caction}
 L\, =  \, F_{\mu\alpha\beta\gamma}F^{\mu\alpha\beta\gamma} \, +  \, C_{\alpha\beta\gamma}\, J^{\alpha\beta\gamma},
\end{equation}
where $F_{\mu\alpha\beta\gamma}\, = \, \partial_{[\mu}C_{\alpha\beta\gamma]}$ is the four-form
field strength, and  $J^{\alpha\beta\gamma}$ is a conserved external current
\begin{equation}
\label{conserved}
\partial_{\alpha}\, J^{\alpha\beta\gamma}\, = \, 0.
\end{equation} 
The action (\ref{Caction}) is  then invariant under the gauge transformation
\begin{equation}
\label{gauge}
C_{\alpha\beta\gamma}  \rightarrow  C_{\alpha\beta\gamma} \, + \, d_{[\alpha}\Omega_{\beta\gamma]},
\end{equation}
where $\Omega$ is a two-form. 
Because of this gauge freedom  in four dimensions $C$ contains no propagating degrees of freedom.
Despite the absence of propagating degrees of freedom,  $C$ nevertheless can create a `Coulomb'-type long-range electric field in the vacuum $ F_{\mu\alpha\beta\gamma} =  F_0\epsilon_{\mu\alpha\beta\gamma}$. In this respect, the $3+1$-dimensional 
three-form gauge theory is very similar to $1+1$ electrodynamics\cite{cjs}. 
  As it is obvious from the equation of motion, in the absence of sources, the four-form electric field can assume an 
{\it arbitrary} constant value.
Its equation of motion
\begin{equation}
\label{fequation}
\partial^{\mu}F_{\mu\nu\alpha\beta} \, = \, 0 
\end{equation} 
is solved by 
\begin{equation}
\label{solution}
F_{\mu\nu\alpha\beta} \, = \, F_0 \,  \epsilon_{\mu\nu\alpha\beta}, 
\end{equation} 
where $F_0$ is an arbitrary constant. 

Thus, the theory has a continuum of the vacuum states each labeled by an expectation value of a constant electric field $F_0$. These vacua obey the super-selection rule.  
 $F_0$ is not a dynamical quantity, and there is no transition between 
the different vacua.  In the other words  no $F$- vacuum is preferred over any other, and any choice of $F_0$ is good. 
In this respect $F_0$-vacua are very similar to the $\theta$-vacua in QCD\cite{theta}.

 As said above,  this connection between the $F$-vacua and QCD $\theta$-vacua is deeper
  than one naively may think, and will play the central role in our approach  to the strong CP problem.  
 The key point in this connection is that  the QCD $\theta$-vacua can be exactly reformulated in terms 
  of the vacua with the four-form electric fields (\ref{gluonfourform}).  We shall come back to this issue shortly. 
  
  Let us now briefly discuss how the superselection rule in the above example gets lifted in the presence of branes.  In the presence of the external source $J^{\alpha\beta\gamma}$ the superselection rule 
 gets partially lifted, permitting transitions between the vacua with certain discretized values of 
 the electric field. 
 As said above, the gauge invariance demands that the three-form be sources by two-dimensional surfaces,
 2-branes, for which the conserved current takes the following form  
 \begin{equation}
\label{current}
J^{\alpha\beta\gamma}(x)\,  = \,  \int d^3\xi \delta^4(x \, - \, Y(\xi))\, q \,  
 \left( {\partial Y^{\alpha} \over \partial \xi^a}
{\partial Y^{\beta} \over \partial \xi^b}{\partial Y^{\gamma} \over \partial \xi^c}\,\right) \epsilon^{abc} ,
\end{equation} 
where $q$ is the charge of the brane, and $x^{\mu} \, = \, Y^{\mu}(\xi)$ specify a $2+1$-dimensional history
of the brane in $3+1$-dimensions as a function  of its world-volume coordinates $\xi^a ~(a=0,1,2)$. 
Obviously, the current $J_{\alpha\beta\gamma}$ is conserved as long as $q$ is a constant. 

 The brane self-action has the  standard form
\begin{equation}
\label{branetension}
 - T\int \, d^3\xi \sqrt{-g},
\end{equation}
where $T$ is the brane tension (a mass per unit surface), and  $g_{ab} = \partial_aY^{\mu}\partial_bY^{\nu}\eta_{\mu\nu}$ is the induced metric on the brane. 
Note that, since the bulk $4$-dimensional gravity plays no essential role in our considerations, we 
have taken a flat Minkowskian  $4$-dimensional metric $\eta_{\mu\nu}$.  
The brane can be taken to be flat and static,  $Y^{\mu} \, = \, \xi^{\mu}$ for $\mu = 0,1,2$, and 
$Y^3 \, = \, 0$. 
The equation of motion then becomes  
\begin{equation}
\label{feqstatic}
\partial_{\mu}\,  F^{\mu\nu\alpha\beta} \, = \, -\, q \delta(z) \epsilon^{\nu\alpha\beta z},
\end{equation} 
where $z=0$ is the location of the brane. 
Equation (\ref{feqstatic}) shows that the brane separates the two vacua 
in either of which  $F_0$ is constant, and the two values differ by $|q|$.
 Thus, the introduction of branes ensures that the transition  between the vacua with different values of $F_0$  is possible, as long as the value of $F_0$ changes by an integer multiple of $q$.
Hence the discrete quantum transition between the different vacua are possible via nucleation of 
closed branes 
 
In the other words, the theory given by the action (\ref{Caction})  has a multiplicity of the
 discrete vacuum states. Among all possible vacua  there are the subsets ( the vacuum families) 
 that are connected via quantum-mechanical tunneling. 
The different vacua within a given family can be labeled by an integer $n$. The value of the field strength in these vacua is  
\begin{equation}
\label{fzeron}
-{1\over 24}F_{\alpha\beta\gamma\mu}\epsilon^{\alpha\beta\gamma\mu} \, =\, F_0 \, = \, qn \, + \, f_0, 
\end{equation}
where $f_0$ is a constant, which is a fixed number for a given family, but changes from family to family.  Thus, within a given family, the 
value of $F$ is quantized in units of the brane charge.

\section{The Strong CP Problem in Three-Form Language}

 As discussed above, the vacua with a constant four-form electric field $F$ are very similar to 
 $\theta$-vacua in CQD. We wish now to show that the real QCD  $\theta$-vacua can be understood as the vacua with different values of an electric field of a composite 
 QCD four-form (\ref{gluonfourform}).  The detailed discussion of this connection can be found in
 \cite{axigauge}. Before going to real QCD let us formulate the $\theta$-vacuum problem in a theory with a free fundamental three-form, with the simplest Lagrangian (\ref{Caction}) and no sources. 
This theory is in the Coulomb phase, and this fact is the source of the generalized strong CP problem. 
As shown above, equations of motion are solved by an arbitrary constant electric field
(\ref{solution}), 
where $F_0$ is arbitrary, and plays the same role as the $\theta$-parameter in QCD.  
In particular, the constant electric field (\ref{solution}) is CP-odd.  Also, 
the $F_0$-vacua obey a superselection rule.  Note that  the expansion 
on a background with a constant electric field $F_0$, generates a direct analog of the 
$\theta$-term
\begin{equation}
\label{thetaf}
\theta \, F,
\end{equation}
where $\theta \, = \, {1 \over 24} F_0$. 
Hence, the $\theta$-parameter in a three-form gauge theory is equivalent to a constant 
four-form electric field in the vacuum.  This is analogous to what happens in the free massless
electrodynamics in two dimensions in which the $\theta$-parameter also appears as an electric field \cite{cjs}.
Thus, the strong CP problem reformulated in the language of a three-form gauge theory, reduces to the following question.
How can the four-form Coulomb electric field be made naturally small?

  The strong CP problem in QCD can be  reformulated in the above-presented thee-form language.  
For this,  consider a $\theta$-term in  $SU(N)$ gauge theory with a strong coupling scale $\Lambda$ (which we shall set equal to one) and no
light fermion flavors 
\begin{eqnarray}
\label{apsi1}
L \, = \, \theta\, {g^2\over 32\pi^2}
 G^a\tilde{G^a}, 
 \end{eqnarray}
where $g$ is the gauge coupling,  and $a$ is an  $SU(N)$-adjoint index.
As discussed in the introduction,   this term can be rewritten as  dual of four-form fields strength $F$ of a composite three-form $C_{\alpha\beta\gamma}$ according to (\ref{apsi}) and (\ref{qcdform}).
Under the gauge transformation, $C_{\alpha\beta\gamma}$ shifts as (\ref{gauge}) with
\begin{equation}
\label{omega}
\Omega_{\alpha\beta}\, = \, A^a_{[\alpha}\partial_{\beta]}\omega^a,
\end{equation} 
where $\omega^a$ are the $SU(N)$ gauge transformation parameters. The four-form field strength
$F_{\mu\alpha\beta\gamma}\, = \, \partial_{[\mu}C_{\alpha\beta\gamma]}$ 
is of course invariant under (\ref{gauge}) and (\ref{omega}). 
Note that the $SU(N)$  Chern-Simons current $K_{\mu}$ can be written as 
\begin{equation}
\label{cs}
K_{\mu} \, = \, \epsilon_{\mu\alpha\beta\gamma} C^{\alpha\beta\gamma}. 
\end{equation} 
It is known \cite{qcdform} that at low energies, the three-form $C$ becomes a massless {\it field}, 
and creates a long-range Coulomb-type constant force.
The easiest way to see that $C$ mediates a long-range interactions is through the 
Kogut-Susskind pole \cite{KS}.   The zero momentum limit of the  following correlator  
\begin{equation}
\label{corr}
 {\rm lim} _ {q \rightarrow 0} \, q^{\mu}q^{\nu} \, \int d^4x  {\rm e}^{iqx} \langle 0| T K_{\mu}(x)K_{\nu}(0)|0\rangle
\end{equation}
is non-zero,  as it is related to topological susceptibility of the vacuum, which is 
a non-zero number in pure gluodynamics.  Hence, the correlator of the two  Chern-Simons
currents has a pole at zero momentum, and the same is true for the correlator of 
two three-forms.  Thus,  the three-form field develops a Coulomb propagator and mediates a  
long-range force.  Because the probe sources for the three-form are two dimensional surfaces 
(domain walls or the 2-branes), the force in question is constant.  

 In the other words,  at low energies, the QCD Lagrangian contains a massless three-form field, and can be written as 
 \begin{equation}
\label{qcdtheta}
L \, = \, \theta F \, + \, {\bf K}(F)\, + \, ...,
\end{equation}
The exact form of the function ${\bf K}$ in QCD is unknown\footnote{Some subleading  terms 
 were estimated in \cite{giga} using the large-$N$ QCD expansion.}, but it is unimportant for our purposes.  
It is obvious now that the $\theta$-problem in  QCD is isomorphic to a problem of a constant 
four-form electric field, and that QCD $\theta$-vacua are simply  vacua with different values of this electric field. 

Notice that the axion solution of this problem is nothing but Higgsing the composite three-form given in 
 (\ref{qcdform}).
That is, the axion solves this problem by giving a gauge-invariant mass to the three-form field and screening its electric field in the vacuum.\footnote{Alternatively one could try screening $\theta$
by dramatically altering the topology of space\cite{shap}}.
 Indeed, the axion solution is based on the idea of promoting 
$\theta$ into a dynamical pseudoscalar field $a$, which  gives us the following Lagrangian
\begin{equation}
\label{aFactionqcd}
L \, =  {f^2 \over 2} \, (\partial_{\mu}a)^2 \,  - \,  a \,  F \, + \,{1 \over 24} {\bf K}(F), 
\end{equation}
where $f$ is the axion decay constant. 
The reader can easily check that the minimum of the axion potential in the above theory 
is always at $F=0$, regardless of the form of the function ${\bf K}(F)$, and that the 
theory contains no massless correletors.  In the other words, the three-form gauge theory 
is in the Higgs phase, due to `eating up' the axion field\footnote{The Higgs effect can be clearly visualized by dualizing the axion to an anti-symmetric two-form field\cite{gigamisha, axigauge}.}
The detailed discussion of this phenomenon
can be found in \cite{axigauge}. 

  We shall now choose a different path for solving the strong CP problem. We shall not introduce
  any massless axions. In our approach the four-form electric field will remain in the Coulomb phase, 
  but we shall explain its smallness by altering the structure of the $\theta$-vacua. 
  In our treatment the vacuum  superselection rule will be lifted, but in such a way that 
  the vacua will  accumulate in a  tiny-$\theta$ region.

\section{Promoting $\theta=0$ into the Vacuum Accumulation Point}

We shall now discuss the dynamics that promotes  $\theta = 0$
(Tr$G\tilde{G} = 0$)  into the {\it vacuum accumulation} point. As a result of this dynamics, 
all but a measure zero set of vacua become  piled up near $\theta=0$. 

We shall achieve this effect by replacing the {\it topological} axion current $\epsilon^{\alpha\beta\gamma\mu}\partial_{\mu} a$ in the second term of the Lagrangian (\ref{aFactionqcd}) by the CP-odd 2-brane current, according to (\ref{replace}). In the other words, 
we trade a massless CP-odd scalar  for a massive (P)CP-odd extended object, the 2-brane. 
\vskip 0.5 cm
 {\bf [} The meaning of a CP-odd brane requires some clarification. In fact, a topological  domain 
wall formed by a P- and CP-even scalar $\chi$ represents a simplest field theoretic example
of a P- and CP-odd 2-brane.  Indeed, let us assume that $\chi$ changes  by $\Delta \chi$ through the wall.  The topological current $J^{\alpha\beta\gamma} \, = \, \epsilon^{\alpha\beta\gamma\mu}\partial_{\mu}\chi$ then has the parity which is opposite to 
the one of $C_{\alpha\beta\gamma}$.  Thus, in such a case, the coupling
$C_{\alpha\beta\gamma} J^{\alpha\beta\gamma}$, is not permitted by P and CP and must be 
accompanied by the odd powers of a P-odd  electric field $F$.  In effective low energy description,
in which the wall thickness is integrated out,  the topological current plays the role of the 2-brane
current  which sources $C_{\alpha\beta\gamma}$ 
\begin{equation}
\label{replace1}
\epsilon^{\alpha\beta\gamma\mu}\partial_{\mu} \chi(x) ~~~ \rightarrow ~~~ 
\int dY^{\alpha\beta\gamma}\delta^4(x -Y).
\end{equation} 
In the view of this, if we wished to limit ourselves to entirely field theoretic realization, with no 
fundamental extended object, we could have simply replaced a massless pseudoscalar axion field
with a  heavy  scalar $\chi$, that forms a topological domain walls.  One can easily construct other 
more involved examples of the CP-odd walls.  However, we wish to keep our treatment maximally general, and without specifying any underlying nature of 2-branes. {\bf ]} 
\vskip 0.5 cm


Up to total derivatives the Lagrangian then becomes 
\begin{equation}
\label{newl}
L \, =  \,{1 \over 24} {\bf K}(F) \, + \, \,C_{\alpha\beta\gamma} J_{(T)}^{\alpha\beta\gamma},
\end{equation}
where $ J_{(T)}^{\alpha\beta\gamma}$ is the transverse part of the brane current (\ref{current}). 
The reason for transversality is as follows. From the assumption that the brane is CP-odd 
it follows that the brane charge $q$ can no longer be a constant,  but should be  
an odd continuous  function of $F$.  For example,
\begin{equation}
\label{qattractor}
q\, \rightarrow \, q(F) \,  \propto \, F^{2n + 1},
\end{equation}
where $n$ is a positive  integer.
This is exactly what we need, since according to the general mechanism of \cite{attractors},   the zero of the electric field $F$ will become  an accumulator in the space of vacua.  We shall demonstrate this explicitly in a moment. But let us first note, that in order to maintain the gauge invariance in the case of a field-dependent 
 $q$, we shall follow the prescription of \cite{attractors} and couple $C_{\alpha\beta\gamma}$  and 
 $J^{\alpha\beta\gamma}$ transversely.  That is, we shall adopt the following coupling 
 \begin{equation}
\label{newcoupling}
 \,C_{\alpha\beta\gamma} J_{(T)}^{\alpha\beta\gamma},
\end{equation}
where $J_{(T)}$ is the transverse part of the current
\begin{equation}
\label{transverse}
 J_{(T)}^{\alpha\beta\gamma}\, =  \, \Pi^{[\alpha}_{\mu}J^{\mu\beta\gamma]}.
\end{equation}
Here 
$\Pi_{\mu\nu}\, = \eta_{\mu\nu} \, -\, {\partial_{\mu}\partial_{\nu} \over \partial^2}$ is the transverse projector.

 For constant $q$, we have  $\partial^{\alpha} J_{\alpha\beta\gamma}\, =\, 0$ and $J_{(T)} \, = \, J$.
Thus, the coupling (\ref{newcoupling}) reduces to the one in (\ref{Caction}).
 This fact accomplishes our goal.
In each given vacuum the expectation value of $F$ is fixed to a constant. 
On the other hand, the change of $q$ from vacuum to vacuum is permitted, because $C_{\alpha\beta\gamma}$ only couples to the transverse part of $J_{\alpha\beta\gamma}$. 
The existence of the attractor point at $F=0$ is guaranteed by the fact that  $J_{(T)} \rightarrow 0$  when $F \rightarrow 0$. 

The coupling (\ref{newcoupling}) is the gauge-invariant generalization of (\ref{Caction}) for the
case of a non-constant charge $q(F)$. 
 Although, the coupling (\ref{newcoupling}) contains a projector, 
as shown in \cite{attractors} it is local, and can be obtained from a  local underlying theory after integrating out the St\"uckelberg field.  For completeness, we repeat this derivation in the appendix. 

 Putting all the ingredients together, let us now show that the theory has an attractor point in the space of vacua at $F\, =\,0$.  Since the existence of the attractor point at $F=0$ is independent on the form
 of the function ${\bf K}(F)$, we shall take ${\bf K}(F) \, = \, F^2/2 $ for simplicity 
\begin{eqnarray}
\label{theaction}
L \, &=& \,  {1 \over 48 } F^2 \, + \,C_{\alpha\beta\gamma} J_{(T)}^{\alpha\beta\gamma}.
\end{eqnarray}
The equations of motion are
\vskip 1cm
\begin{equation}
\label{cequation}
\epsilon^{\mu\alpha\beta\gamma} \partial_{\mu} \left (F \, + \,24 C_{\kappa\rho\tau} 
{\partial J_{(T)}^{\kappa\rho\tau}  \over \partial F}\right)\, = \, J_{(T)}^{\alpha\beta\gamma},
\end{equation}
\vskip 1 cm
The vacuum structure of a similar toy example of $1+1$-dimensional electrodynamics, was studied in  
\cite{toy}.  The analysis in our case is pretty similar. 
In order to visualize the vacuum structure, we have to figure out how the constant four-form 
electric field $F$ changes at the static brane.  
For definiteness, we shall place the latter at 
$x^3 = z = 0$. 
Then the only non-zero components of the current become $J^{\alpha\beta\gamma}\, =\, 
\epsilon^{\alpha\beta\gamma 3}\, \delta(z)$. 
We shall now look for the static $z$-dependent configuration $F \, = \, F(z)$ 
and $C^{\alpha\beta\gamma} = \epsilon^{\alpha\beta\gamma 3} \, C(z)$. 
Then the only non-trivial equation is 
\begin{equation}
\label{onlyeq}
d_z\left (F(z) \, + \, 864 \delta(z) C(z) {\partial q(F) \over F} \right ) \, = \, - q(F(z))\delta(z).
\end{equation}
To find a jump in the value of $F$, we integrate the equation in a small interval near $z=0$. 
This integration gives 
\vskip 0.7 cm
\begin{equation}
\label{delta}
\Delta F \, = \, - q(F(0)).
\end{equation}
\vskip 0.7 cm
Because, $q(F)$ is an odd function of  $F$, 
the equation (\ref{delta}) shows that jumps are becoming smaller and smaller as $F \rightarrow 0$, which proves that there are infinite number of vacua within an arbitrarily small neighborhood
of $F = 0$. Thus,  $F=0$ ($\theta = 0$) is a vacuum accumulator point.  

 As it was shown in \cite{attractors},\cite{toy} the number of vacua  with the electric field 
exceeding a certain value $F_*$  diverges with $F_* \, \rightarrow \, 0$ as 
 \begin{equation}
\label{number}
n_{F_*} \, \sim \ \int_{F_*}^1 \, {dF\over q(F)}.
\end{equation}
For $q(F) \, = \, F$, this  gives 
\begin{equation}
\label{numberf}
n_{F_*} \sim {\rm ln} \left ({1\over F_*}\right )
\end{equation}
Translated to the statistics of the $\theta$-vacua, this implies that the number of $\theta$ vacua
within each family diverges as 
\begin{equation}
\label{ntheta}
n_{\theta} \sim {\rm ln} (\theta^{-1})
\end{equation} 
for $\theta\, \rightarrow \,  0$.

\section{The Accumator-Shift Problem and the Fix}

 We shall now address a potential problem which could destabilize the above solution of the strong CP 
 problem. The problem may arise from the perturbative gluon loops that may generate a constant 
 part in $q(F)$ and therefore shift the accumulator point  away from $F=0$ ($\theta =0$).  Such a constant part may be
 generated because of the following reason.  The symmetries that guarantee that $q(F)$ is an odd function of $F$ are P and CP. However,  both symmetries are  broken in the 
electroweak sector of the Standard Model, and this breaking will result into a contribution  
to the $\theta$-term, which we shall call $\theta_{SM}$. We shall assume that $\theta_{SM} \, \sim \, 1$.
 The breaking of CP by the $\theta_{SM}$-term can (and in general will) be communicated to 
 the brane via the gluon loops, and this will result into the shift of the function $q(F)$ by an 
 $F$-independent constant proportional to the $\theta_{SM}$-term 
 \begin{equation}
\label{shift}
q(F) \rightarrow q(F) \, + \, \theta_{SM} \Lambda_{c}^4, 
\end{equation}
 where $\Lambda_{c}$ is a cutoff. 
Such a shift by a large constant would be a disaster for our solution, since the vacuum accumulation point 
 now would be shifted to $F \neq 0$ value, that in general would result into 
 $\theta_{observable} \sim \theta_{SM}$. 

 We shall now solve the above problem dynamically.  For this we shall assume that 
 CP is broken spontaneously at some scale $M$, by a
 non-zero VEV of some scalar field  $\phi$. In such a case all the CP-violating operators, and in particular the $\theta_{MS}$-term will be generated from some effective high-dimensional operator of the form 
 \begin{equation}
\label{phiggdual}
\left ({\phi \over M}\right )^N\, {\rm Tr} G\tilde G \, \rightarrow \, \theta_{SM} F,
\end{equation}
where $N$ is some power, which depends on the details of model building and in particular on 
quantum numbers of the field $\phi$ under different gauge symmetries.  After $\phi$ gets a VEV, the above coupling 
will translate into an effective $\theta_{SM}$-term
with  $\theta_{SM}  = \left ({\langle \phi\rangle \over M}\right )^N$. 

 Notice that, we are {\it not}  demanding
any particular smallness of $\theta_{SM}$. Because in the vacuum $\langle \phi \rangle \sim M$, 
the $\theta_{SM}$ is naturally of order one.  So in the absence of our CP-odd branes the strong CP problem would be there, as usual.  What we gain by promoting $\theta_{SM}$ into a VEV of a field, 
is that its expectation value can diminish in the vicinity of the brane due to the influence of the latter. This fact  solves the accumulator shift problem.  The shift of the brane charge by the gluonic loops
now will manifest itself  through correcting $q(F)$ by the following local operator 
 \begin{equation}
\label{shift}
q(F) \rightarrow q(F) \, + \,  \left ({\phi  \over M}\right )^N\Lambda_c^4. 
\end{equation}
The important point is that in evaluating this shift,  the value of $\phi$ has to be taken not in the vacuum, but at the location of the brane. Due to the influence of the brane the latter value can be negligibly small, as we show in a moment. 
In this way the accumulator shift  is avoided. 
  
The suppression of the  $\phi$ VEV on the brane can happen, because $\phi$ field is allowed to have various potential terms on the brane world volume, compatible 
with symmetries. The most important of these is a brane-localized mass term, which can be
introduced in the four-dimensional action in the following form
\begin{equation}
\label{massp}
-\,{1 \over 2} \int \, dx^4 \,  M_{br}(x)^2\phi^2,
\end{equation}
where 
\begin{equation}
\label{branemassterm}
M_{br}^2(x)\, =\, \pm \int\, d\xi^3\sqrt{-g}\, M_B \delta^4(x-Y).
\end{equation}
In the above expression $M_B$ is a positive mass parameter.  As shown in \cite{attractors} in the case of the positive sign, the brane localized mass term has an effect of suppressing the $\phi$ VEV on the brane.  The source for this suppression is easy to understand. 
The equation for $\phi$ in the  background of the brane (located at $z=0$) is 
\begin{equation}
\label{eqbrane}
\partial^2 \phi\, - \,  \lambda^2\, (\phi ^2 \, -  m^2)\,\phi \, - \, \delta(z)M_B\,\phi \, = \,0.  
\end{equation}
Here we have assumed that the bulk potential for $\phi$ field 
is $V(\phi) \,= \, {\lambda^2\over 4} (\phi^2 \, - \, m^2)^2$ so that the bulk VEV is $\phi_{bulk} \, = \, m$.
 From (\ref{eqbrane})  it is clear that the positive brane-localized mass term is seen by 
the field as a potential barries, which  for $M_B \gg \lambda m$ strongly suppresses  the expectation value on the brane, $\phi_{br} \ll \phi_{bulk}$.  

 $\phi_{br}$ can be estimated by minimizing the following expression
(we ignore the factors of order one)
\begin{equation}
\label{energymin}
E \, =\, M_B\phi_{br}^2 \, + \, (\phi_{br} - m)^2m\, + \, \lambda^2 \, (\phi_{br}^2\, - \, m^2)^2m^{-1}
\end{equation}
The first term in this expression comes from the brane mass term. The second and the third terms are
the expenses in the gradient and the bulk potential energies.
The full expression is minimized at 
\begin{equation}
\label{p0min}
\phi_{br} \sim \phi_{bulk} {m_{\phi} \over M_B},
\end{equation}
where $m_{\phi} \, \equiv \, \lambda m$ is the bulk mass. 
Thus, in any given vacuum, the brane expectation value of $\phi$ is 
by the factor ${m_{\phi} \over M_B}$ smaller relative to its bulk counterpart.

This statement can be checked exactly.  The equation  (\ref{eqbrane}) has an explicit solution 
  \begin{equation}
\label{exactsol}
\phi(z) \, = \, m \, {\rm th} \left [{m_{\phi} \over \sqrt{2} }\left (|z| \, +\, {1 \over \sqrt{2} m_{\phi}} {\rm arcsh}({2\sqrt{2} m_{\phi}\over M_B})\right ) \right ], 
\end{equation}
 which  confirms the above estimate, since  $\phi_{br} \, =\, \phi(0) \,\simeq \, \sqrt{2} mm_{\phi}/M_B$.

To summarize, by decreasing the bulk mass of $\phi$ relative to its brane mass,  the expectation
 value at the brane gets diminished. The brane surrounds itself by a `halo'  of a restored CP
 region.  The size of this halo is $\sim m_{\phi}^{-1}$.  Inside this region, the only source of the 
 CP violation is the expectation value of $F$ that is sources by the CP-odd brane.

  Due to the above effect,  the accumulator gets shifted to the point 
 \begin{equation}
\label{newnvacua}
q(F)  \sim \left({m_{\phi}\over M_B}\right)^{N}. 
\end{equation}  
This value can be naturally extremely small.  To get a rough feeling about the possible smallness, 
consider an extreme case, when the brane has a Planck scale tension. Then, we can take $M_B\sim 
M_P$.  If $\phi$ is a modulus that couples to the Standard Model  fields via $M_P$ suppressed interaction,  the natural lower value 
for $m_{\phi}$ is somewhere around $10^{-3}$eV. This will be the case if the supersymmetry breaking scale is around TeV.  Then taking  $q(F) = F$ the value of the observable
$\theta$-term at the accumulator point will be $\theta_{observable} \sim 10^{-31N}$.  Which is practically
unobservable.  Phenomenologically more interesting situation occurs for the lower values of the brane tensions, in  which case predicted $\theta_{observable}$ may be close to the phenomenological lower bound, 
and be potentially observable.

\section{Experimental Signatures}

We wish to briefly discuss some possible experimental signatures of the presented scenario. 
The crucial role is played by the branes to which the Standard Model gauge fields are coupled. 
These branes can be introduced either as some field theoretic domain walls (see \cite{attractors}), or as the fundamental 
objects. The latter possibility is phenomenologically the most interesting one, since branes 
can be produced at particle colliders. 

 There is no obvious particle physics upper bound on the tension of these branes  (although
the specific cosmological considerations may place one).  In this respect, the role played 
by the brane tension in our scenario is analogous to the one played by the axion decay constant
in the Peccei-Quinn solution. The latter can be arbitrarily high,  allowing axion to be arbitrarily weakly coupled and practically unobservable\cite{invisible}. 

 However, if the brane tension is at the TeV scale or lower, the branes may be observed at LHC in form of the resonances with a specific spectrum. 
  One can think of number of possible production channels
for these resonances.  Because branes are predicted to be coupled to
 all possible CP-odd combinations of the Standard Model Chern-Simons forms and dual field-strengths, 
 they can be produced in gauge boson scattering.
For instance, they can be produced in gluon-gluon collision
accompanied by the production of the additional gauge fields. A typical process would be 
\begin{equation}
\label{production}
2 \, g \, \rightarrow \, {\rm brane~resonance} \, + \, 2\,\gamma, 
\end{equation}
in which two gluons produce a brane resonance and the two photons.  The brane resonances can then decay into a combination of the standard model gauge fields (for instance,  into four photons
(gluons, weak bosons)). They can also decay into the CP-violating scalar $\phi$, which then (if mass allows) can decay 
into the Standard Model fermion-antifermion pair and a Higgs. 

 Note that the brane tension may be below the $W,Z$ masses, in which case the decay in this particles is excluded. Also note that the mass of $\phi$ in the extreme case can naturally be as small as $10^{-3}$eV. This will be the case if supersymmetry breaking scale is TeV, and $\phi$ couples via 
 Planck scale suppressed operators to the Standard Model fields.  In such a case $\phi$ 
 can only decay into the photons, and the lifetime will be too long for being observed within the detector. 
In such a case, $\phi$ can also manifest itself through a new gravity-competing force at sub-millimeter 
scales.  

Brane resonances can also be produced in quark-antiquark annihilation together with the 
 Higgs in the final state, via intermediate $\phi$ scalar. 

 The generic distinctive feature of brane resonances will 
 be their characteristic spectrum.  For instance, it is known\cite{membrane} that the radial excitations of a spherical membrane have the spectrum of a radial Schr\"odinger equation with a quartic  potential
 (for simplicity we work in units of the brane tension)
  \begin{equation}
\label{spectrum}
\left (-{d^2 \over dr^2} \, + \, r^4 \right)\, \psi(r)  \, = \, m^2\psi(r),  
\end{equation}
which has a $m^2 \sim n^{4/3}$ scaling for large $n$.

\section{Conclusions}

The source of the strong CP problem is that the QCD has a continuum of the vacuum states 
labeled by the parameter $\theta$, which obey the superselection rule and most of which
are excluded observationally.  The phenomenologically acceptable values $\theta < 
10^{-9}$ are not preferred by any statistics or dynamical considerations. 

 What we have achieved in our treatment is that we have lifted the superselection rule in such a way that 
 the transition within an infinite subset of vacua,  called a {\it  vacuum family}, is now permitted by 
 nucleation of  branes.   A given family of vacua can be constructed by randomly choosing the value of the 
 parameter $\theta$ and adding all possible vacua that can be created from it by nucleation of the arbitrary number of branes.  The set of vacua created in this way will by default be isolated from the other sets by the superselection rule. Thus, the transition among the families is still forbidden.  A priory, the number of families may be large or even infinite, but within each family the accumulation of the infinite 
 number of vacua happens within the phenomenologically permitted region $\theta < 10^{-9}$. 
 
  In this way, in the full theory, the  phenomenologically unacceptable vacua forms a measure-zero set.  
 
In this work we were mainly concerned with the vacuum statistics, and  we did not discuss an
explicit  cosmological scenario that in real time picture would drive the Universe towards the vacuum accumulation point.  However, it is intuitively clear that in any early Universe scenario in which the 
statistics of vacua matters, the highest probability will be given to the phenomenologically acceptable 
ones, due to their enormous number.

 We also wish to remark that, although fundamentally different, the presented mechanism shares some spiritual connection with the irrational axion idea\cite{irax}. 

 Finally, there are possible experimental signatures of the above theory.
The general prediction is the existence of  the 2-branes to which the standard model gauge fields are coupled.  The tension of these branes  may be around TeV scale or lower. In such a case, they can be
produced in particle collisions in form of the resonances  with a characteristic  spectrum.

\vspace{0.5cm}   

{\bf Acknowledgments}
\vspace{0.1cm} \\
 We thank Gregory Gabadadze and Mikhail Shaposhnikov  for  discussions, and to
 Roman Jackiw for  discussions on the related issues.  We thank Ann Nelson
 and Neal Weiner for comments.  
This work is supported in part  by David and Lucile  Packard Foundation Fellowship for  Science and Engineering, and by NSF grant  PHY-0245068

\section{Appendix} 

 In this appendix, we briefly repeat the method of ref\cite{attractors} for obtaining the 
 couplings (\ref{transverse}) from the local gauge-invariant theory.  This is achieved by introducing the 
following couplings in the Lagrangian
\begin{equation}
\label{thetacoupling}
(C_{\alpha\beta\gamma} \, - \, \partial_{[\alpha} B_{\beta\gamma]} )\, J^{\alpha\beta\gamma} \, + \, 
X^{\beta\gamma} \partial^{\alpha} (C_{\alpha\beta\gamma} \, - \, \partial_{[\alpha}B_{\beta\gamma]}).
\end{equation}
Here, two-form $B_{\alpha\beta}$ is a compensating  St\"uckelberg field,  which under the gauge transformation (\ref{gauge})
shifts in the following way
\begin{equation}
\label{gaugeB}
B_{\alpha\beta} \, \rightarrow  \, B_{\alpha\beta}  \, + \, \Omega_{\alpha\beta}.
\end{equation}
$X^{\beta\gamma}$ is a two-form Lagrange multiplier that through its equation of motion
imposes the transversality constraint 
\begin{equation}
\label{Beq}
\partial^{\mu} \partial_{[\mu}B_{\alpha\beta]} \, = \, 
\partial^{\nu} C_{\nu\alpha\beta}. 
\end{equation}
Integrating out the $X$- and $B$-fields,  we 
arrive to the effective coupling (\ref{transverse}).

 Let us briefly comment on a potential effect, discussed in\cite{attractors}, 
 which is a possible screening of the electric field $F$ by the virtual brane loops. 
 The effect is somewhat analogous in spirit to the charge screening by fermion 
loops in the massless  Schwinger model\cite{screening}, except that this issue in the brane case 
is more subtle.
 The point is that the loops of the branes could in principle generate operators of the form
\begin{equation}
\label{CR}
C^{\mu\alpha\beta} \, R \,\Pi_{[\mu}^{\nu}
C_{\nu\alpha\beta]}
\end{equation}
where $R$ is some function that can be expanded in series of the positive powers of  $\partial^2/M^2$, where $M$ is 
a cut-off.  If this expansion contains a constant $\partial^2$-independent term, the propagator of the 
three-form would acquire a physical pole. This would screen the electric field $F$.   In this case the 
strong CP problem would be solved automatically, just as in the axion case, and there would be no need to allude to the vacuum accumulation effects.


\begin{thebibliography}{99}


\bibitem{theta}
C.G.~Callan, R.F.~Dashen and D.J.~Gross,  Phys. Lett. B63 (1976) 334;

R.~Jackiw and C.~Rebbi,  Phys. Rev. Lett. 37 (1976) 172. 

\bibitem{review}  For a review see, J.E. ~Kim, Phys. Rep. 150 (1987) 1.  


\bibitem{qcdform} M.~L\"uscher, Phys. Lett. B78 (1978) 465. 

\bibitem{axigauge} G.~Dvali, hep-th/0507215.

\bibitem{pq} 
 R.D.~Peccei and H.R.~Quinn, Phys. Rev. Lett. 38 (1977) 1440;
Phys. Rev. D16  (1977) 1791.

\bibitem{axion} 
S.~ Weinberg, Phys. Rev. Lett. 40 (1978) 223;

F.~Wilczek, Phys. Rev. Lett. 40 (1978) 279

\bibitem{gigamisha} G.~ Gabadadze and M.~Shifman, Phys. Rev. D61 (2000) 075014, 
hep-th/ 9910050.


\bibitem{vw}
C.~Vafa and E.~Witten,  Phys. Rev. Lett. 53 (1984) 535.

\bibitem{giaalex}
G.~Dvali and A.~Vilenkin,   Phys. Rev. D70 (2004) 63501, hep-th/0304043.

\bibitem{attractors}  G.~Dvali, hep-th/0410286.


\bibitem{cjs} S.~Coleman, R.~Jackiw, L.~Susskind,  Ann. Phys. 93 (1975) 257.



\bibitem{KS} J.~Kogut and L.~Susskind, Phys. Rev. D11 (1975) 3594.



\bibitem{giga} G.~Gabadadze, Nucl. Phys. B552 (1999) 194, hep-th/9902191.


\bibitem{shap} S.~Klebnikov and M.~Shaposhnikov,  Phys. Rev. D71 (2005) 104024, hep-th/0412306


\bibitem{toy} A.~ Nunez and S.~Solganik, hep-th/0506201.


\bibitem{invisible}
 J.E. ~ Kim, Phys. Rev. Lett. 43 (1979) 103;

M.A.~Shifman,  A.I.~Vainshtein and V.I.~Zakharov, Nucl. Phys. B147 (1979) 385.


 A.R.~Zhitnitsky, Sov. J. Nucl. Phys. 31 (1980) 260;

M.~Dine, W.~Fischler and M.~Srednicki, Phys. Lett. B104 (1981) 199.


\bibitem{membrane} 

P.A.~Collins and R.W.~Tucker, Nucl. Phys. B112 (1976) 150; 

M.~L\"ucher, Nucl. Phys. B219 (1983) 233;

B.~Simon, Ann. Phys. 146 (1983) 209. 

A phenomenological application for modeling the glueball spectrum,  was discussed in 
G.~Gabadadze,  Phys. Rev. D58 (1998) 094015, hep-ph/9710402.

\bibitem{screening}
D.J.~ Gross, I.R.~Klebanov, A.V.~Matytsin and A.V.~Smilga,  Nucl.Phys. B461 (1996) 109, hep-th/9511104.

\bibitem{irax}  T.~ Banks, M.~ Dine and N.~Seiberg, Phys. Lett. B273 (1991) 105, hep-th/9109040.


\end{thebibliography}
\end{document}